\documentclass[journal]{IEEEtran}
\usepackage{cite}
\usepackage{amsmath,amssymb,amsfonts}
\usepackage{algorithmic}
\usepackage{graphicx}
\usepackage{mathtools}
\usepackage{color,soul}
\usepackage{epstopdf}
\usepackage{graphics}
\usepackage{amsmath}
\usepackage{algorithm}
\usepackage{amsfonts,amssymb,amsthm}
\usepackage{upgreek}
\usepackage[tight]{subfigure}
\usepackage{multirow}
\usepackage{tabularx}
\usepackage{dblfloatfix}    
\usepackage[normalem]{ulem}

\usepackage{acronym}

\acrodef{RIS}{Reconfigurable Intelligent Surface}
\acrodef{LIS}{Large Intelligent Surface}
\acrodef{SDM}{Software-Defined Metamaterial}

\begin{document}
\title{Fabrication and characterization of Fused Deposition Modeling 3D printed mm-scaled metasurface units}
\author{Anna C. Tasolamprou,
Despoina Mentzaki, 
Zacharias. Viskadourakis,
Eleftherios N. Economou, 
Maria Kafesaki 
and George Kenanakis
\thanks {Anna C. Tasolamprou, Despoina Mentzaki, Zacharias Viskadourakis, Eleftherios N. Economou, Maria Kafesaki and George Kenanakis are with the Institute of Electronic Structure and Laser, Foundation for Research and Technology Hellas,  Heraklion 70013, Greece,  \emph{(Corresponding author: Anna C. Tasolamprou, email: atasolam@iesl.forth.gr)}}}

\maketitle
\begin{abstract}
We present a cost-effective, eco-friendly and accessible method for fabricating three-dimensional, ultralight and flexible millimeter-scale metasurfaces using a household 3D printer. In particular, we fabricate conductive Spilt Ring Resonators (SRRs) in a free-standing form, employing the so-called Fused Deposition Modeling 3D printing technique. We experimentally characterize the samples through transmission measurements in standard rectangular waveguide configurations. The structures exhibit well defined resonant features dependent on the geometrical parameters and the infiltrating dielectric materials. The demonstrated 3D printed components are suitable for practical real-life applications while the method holds the additional advantage of the ecological approach, the low cost, the flexibility and the small weight of the components.  Thus, the flexible and light 3D printed metasurfaces may serve as electromagnetic components and fabrics for coating a plethora of devices and infrastructure units of different shapes and size.  
\end{abstract}

Metamaterials and their two dimensional analogue, metasurfaces, are artificial materials with purposefully designed subwavelength, periodic elementary units, the meta-atoms, which provide controllable interaction with electromagnetic (EM) waves and enable exotic electromagnetic functions.~\cite{Soukoulis2011} Metamaterials have been extensively experimentally evaluated and tested for real life applications and devices, such as antennas, sensors, splitters, modulators, electromagnetic shieldings, energy harvesters~\cite{Tasolamprou2014,Tasolamprou201513972,Glybovski2016,Hintermayr2019,Tasolamprou2017b,Xu2019,Assimonis2019}. Their properties depend mainly on the meta-atoms architecture. One of the most well-known and widely studied meta-atoms is the Split Ring Resonator (SRR). SRRs consist of metallic loops with gaps, and have been studied throughout the years in several variants.~\cite{Soukoulis2007,Gundogdu2007} 

For microwave appplications, meta-atoms dimensions lay in the range of mm and are conventionally fabricated on printed circuit boards (PCBs). \cite{Smith2004,Gundogdu2007} Typical PCB-based fabrication includes methods like chemical etching and computer  control  milling. Despite their advantages, like the production of high precision parts, those methods limit the design space to rigid, two dimensional, planar structures, they involve the use of lossy  substrates such as FR-4 with substantial power absorption, and the handling of toxic etchants. Lately, there has been a great interest in the construction of complex objects employing 3D printing technologies. 3D printing is an additive manufacturing procedure, where a complex 3D structure/object is constructed, by stacking material layers fully controlled by a  computer assisted design (CAD) file.  3D printing process exhibits remarkable advantages, i.e., it is quick, cost-effective and user-friendly / eco-friendly  (there is no need of clean rooms and handling chemicals and reagents) and it exhibits high resolution printing aspects. 

To date several 3D printing methods have been developed, namely stereolithography, selective laser sintering, digital light processing, binder printing, inkjet printing, laminate object manufacturing~\cite{Gebhardt2016,Papadopoulos2018}. Additionally there is the Fused Deposition Modeling (FDM), an effective 3D printing method, using thermoplastic materials in the form of long wires, called filaments. Filaments are heated above their melting point, and then they are extruded through a narrow nozzle. The nozzle builds the anticipated pattern by moving in all $xyz$ directions, controlled by a computer through a corresponding CAD file. The nozzle builds the anticipated pattern, layer by layer, by extruding the molten filament. FDM process includes all the 3D printing advantages mentioned above. Moreover, FDM printers with exceptional printing characteristics (high printing accuracy, option of multiple printing filaments, heating sample beds, etc.) are commercially available, in reasonable prices, enabling the wide use of 3D printing technology, instead of other  well established development methods. 

Considering the above, the employment of the eco-friendly, low-cost FDM method in the fabrication of flexible (no rigid substrate is necessary), fabric-like metasurfaces for microwave applications appears to be a challenging, but  promising idea. The challenge lies in the filament materials. Metamaterials suitable for microwave applications involve metallic components, while most of the available thermoplastic materials for the FDM are fully insulating or have quite law conductivity values. Fabrication of metallic-like metamaterial components via FDM is a not straightforward task and requires either additional fabrication steps for the metallization or careful selection of the  designs and the FDM filaments. 

In this context, we hereby present the fabrication and characterization of free-standing rectangular conductive SRRs, by employing the FDM technique. For the fabrication we use two different types of filaments resulting in the production of two separate SRR series. One of them is made using polylactic acid (PLA) filament, a widely known polymer material. Since PLA is insulating, the constructed SRR pieces, are painted over with conductive silver epoxy. The other SRR series is built using a commercially available filament known as Electrify. This filament consists of a Polyvinylidene chloride (PVDC) matrix, in which copper (Cu) nanoparticles are included in a weight ratio 20\%w/w and therefore the filament exhibits significant electrical transport properties. Each SRR series consists of several species with varying geometry. Both SRR series are systematically characterized experimentally with the use of standard rectangular waveguides while the measurements are corroborated with numerical simulations. Results concretely show that the 3D printed SRR structures exhibit electromagnetic resonance with characteristics similar to those obtained for traditional PCB-based SRRs. Additionally we evaluate the tunability and sensitivity of the structures response to infiltrating dielectric material. Results show that the 3D printed SRRs exhibit a sizable resonance shift, with respect to moderate dielectric constant changes. Overall, we demonstrate that 3D printed SRR structures could be plausibly used in potential microwave applications.

\begin{figure}
\includegraphics{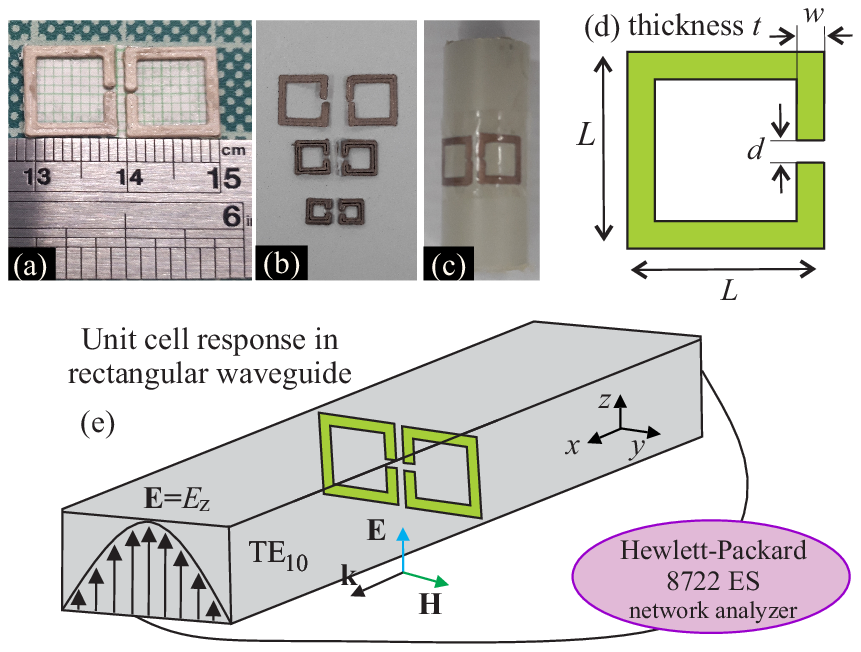}
\caption{\label{fig1} (a) Units of PLA coated with conductive Ag epoxy SRRs with opposing gaps attached on the top of a millimeter page. (b) SRR configurations of various dimensions made of Electrifi. (c) Electrifi SRRs (10 mm) attached to a plastic cylinder of 16 mm diameter. (d) Schematic of the SRRs unit and dimensions. (e) Characterization set-up;  the meta-atoms are placed in the middle of the waveguide and the polarization of the field is along the side of the gaps.  }

\end{figure}
We investigate the rectangular SRRs  pictured in Fig.~\ref{fig1}. The structures were designed using “Tinkercad” (free online 3D design and 3D printing software from Autodesk Inc.). The final design was transformed into a CAD-like file and imported in the 3D printer to build the SRRs.   The structure consists of two SRRs oriented so that their gaps are next to each other as shown in Fig.~\ref{fig1}.  Such SRR geometry exhibits specific advantages, i.e. it is simple to fabricate via FDM 3D printing technique and leads to increased enhancement of the electromagnetic field confinement which results in the high tunability of the  response  with relatively small changes in its environment.
\cite{Rao2018,Chowdhury2011,Penciu2008}  We choose to fabricate structures of variable length $L$, width $w$, thickness $t$ and size of gap $d$ [see Fig.~\ref{fig1}(d)]. The distance between the adjacent SRRs is equal to the wire thickness $w$. 
The components were fabricated by employing the FDM printing procedure. In particular, for the purposes of the current study, we used a Makerbot Replicator 2x, 3D printer and  the two different, commercially available  filaments  as printing materials, i.e., the polylactic acid (PLA) and the Electrifi. Notably the Electrifi has been already used in fabrication of microwave metamaterials. \cite{Xie2017}  Since the filaments are  different materials, different printing conditions should be used for optimum printing results as shown in Table 1.

\begin{table}
\begin{center}

\caption{\label{tab:table1}Printing conditions for
PLA and Electricfy filaments }

\begin{tabular}{lcr}
Printing Conditions&PLA&Electrify\\
\hline
Filament Diameter & 1.75 mm  & 1.75 mm\\
Filament Conductivity & 0 & $10^4 $ S/m\\
Nozzle Diameter & 0.4 mm & 0.4 mm\\
Nozzle Temperature & $230^\circ$C & $140^\circ$C\\
Printing bed temperature& 	$110^\circ$C	&Room temperature\\
Printing speed& 	25 mm/sec&	15 mm/sec
\end{tabular}
    
\end{center}
\end{table}

\begin{figure}
\includegraphics{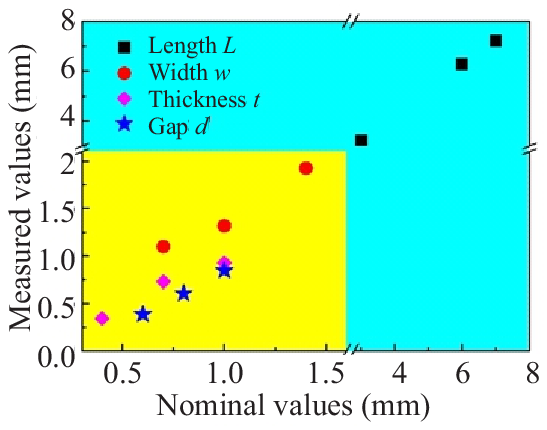}
\caption{\label{fig2} Measured  versus nominal dimensions  for PLA/Ag epoxy 3D printed SRRs. }
\end{figure}

Two different series of SRRs were constructed for each filament. We  covered the PLA-built  SRRs with a thin layer of (a commercially available) conductive silver epoxy (conductivity $\sigma \sim 10^5$ S/m). The obtained SRRs consist of an insulating PLA core, which is encapsulated into a  shell  of a metallic layer.  Circular, Ag-plated SRR structures, have been already grown and studied regarding their microwave metamaterial properties, \cite{Ishikawa2017} rendering the metal-coated, polymer-based SRR a promising design for microwave applications.  We measured their physical dimensions and compared them with the initial CAD file. As seen in Fig.~\ref{fig2}, we found that the real length $L$, is constantly larger than the CAD value, by $\sim$0.2 mm. The maximum difference between the two values is less than 8\%. Similarly the measured thickness of the SRRs is less than the corresponding nominal values, by $\sim$0.03 – 0.07 mm, thus the maximum difference is less than 14\%.  In contrast, the width $w$ is always lower than it nominal value by $\sim$0.3 - 0.5 mm, thus larger difference is observed (up to 55\%).  The real gap is $\sim$0.2 mm narrower than the drawn one. Furthermore, visual inspection of the gap (Fig.~\ref{fig1}) shows that the gap walls are not sharp-ended, but rather round. The rather high differences between nominal and real dimensions along with the not well-shaped gap, imply a sizable low printing resolution. Although the printing resolution of the used printer is $\sim$100 $\mu$m  according to the manufacturer, in our study the printing resolution is estimated around $\sim$ 0.3 – 0.4 mm. In general, the printing resolution is affected by the nozzle diameter, the nozzle speed, the extrusion speed, as well as the filament type. In the current study all the above parameters have been chosen according to the PLA filament manufacturer guidelines. However, the rather low dimensions of the sample ($\sim$1 mm), in combination with the nozzle diameter (0.4 mm) probably contribute contradictorily towards the suppression of the final printing resolution. Regarding the printing quality, the printed SRRs are smooth, surface-clean, and robust (Fig.~\ref{fig1}).  The quality can be further improved by increasing the printing resolution. Furthermore, the samples are quite flexible, as shown in Fig. 1e, in which 10 mm SRRs are attached to a cylinder of diameter 16 mm. Similar conclusions are extracted regarding the Electrifi SRRs.

The SRR components were placed on top of a thin piece of paper (see Fig. 1), which is totally transparent in the microwave regime,  hence, they are considered to be electromagnetically free-standing. The  electromagnetic response of each SRR pair is measured with the use of a standard rectangular waveguide [Fig.~\ref{fig1}(e)]. In particular, we perform transmission measurements using a Hewlett-Packard 8722 ES vector network analyzer. For the characterization  we use waveguides of variable sizes with single mode frequency of operation that in total covers the range 3- 16 GHz; in particular we use the WR187-137-90-62 waveguides. In the single mode operation the waveguide propagation mode is polarized along the small side of the rectangular cross-section as seen in Fig. Fig.~\ref{fig1}(e) and in detail explained in the Supplementary Information. The SRR units under investigation are placed in the middle of the waveguide. Measuring the units in this way provides  a ready, closed-system, characterization of the electromagnetic response.

\begin{figure}
\includegraphics{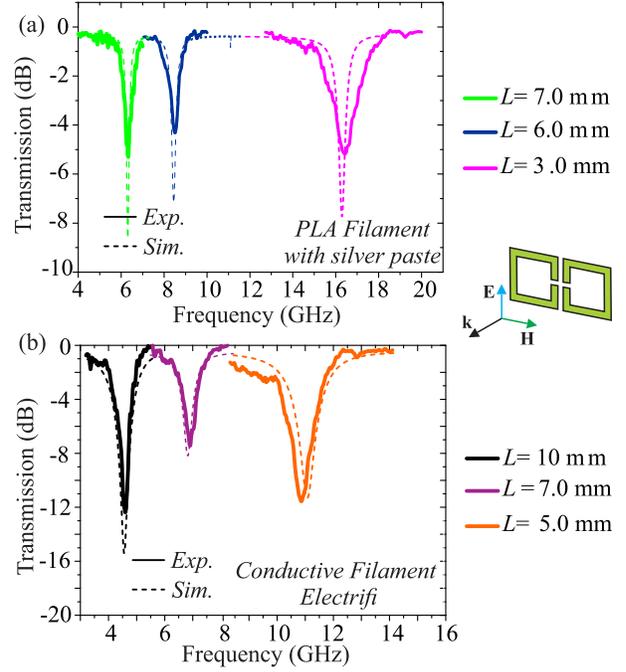}
\caption{\label{fig3}   Electromagnetic investigation of the (a) PLA / Ag epoxy and (b) electrifi built SRRs structure, experimental (solid curves) and  numerical (dashed curves) transmission spectra.  Cases of structures with variable length, $L$ = 3-10 mm  are presented. For all cases, the geometrical parameters are $w$ = 1 mm, $t$ = 0.4 mm and $d$ = 0.6 mm as shown in Fig.~\ref{fig1}. }
\end{figure}

Figure~\ref{fig3}(a) shows the experimental evaluation of the PLA/Ag epoxy SRR samples along with the numerical corroboration. In particular, we measured three unit cells with $L= 3,6,7$~mm. All SRR structures show well-defined, moderately sharp (-4 to -5~dB) resonances. The first thing to comment is the level of the transmission dip which is particularly shallow with respect to conventionally built metasurfaces \cite{Katsarakis2007}. This is a direct consequence of the produced material and the rectangular waveguide characterization.  The decreased electrical conductivity exhibited by the 3D-printed PLA / Ag epoxy SRRs, in comparison to the PCB-built ones  leads to the inductance of weaker currents in the meta-atoms. The  conductivity decrease directly results to reduction of the quality factors of the resonances, that is, the amplitude resonance transmission, along with a resonance broadening also accompanied by a shift to lower frequencies. Relative numerical study is presented in the Supplementary Information. Additionally the relative poor amplitude of the resonance is connected with the characterization of the sample in the waveguide which effectively involves weaker intercoupling of the meta-atoms as discussed in the Supplementary Information. As expected, the largest SRR pair shows resonance at $\sim$3 GHz, while the smallest pair shows resonance at $\sim$16 GHz, a result of the structures scaling. The numerical simulations  successfully reproduce the resonance frequency, for all structures. Some discrepancies are the result of various parameters such as the low-resolution of the 3D printing process, as discussed previously, the application of the silver paste which can result in non-uniformly thick shells with effectively lower conductivity, possible deformations and the details of the electromagnetic characterization in the rectangular waveguides. 

\begin{figure*}
\includegraphics{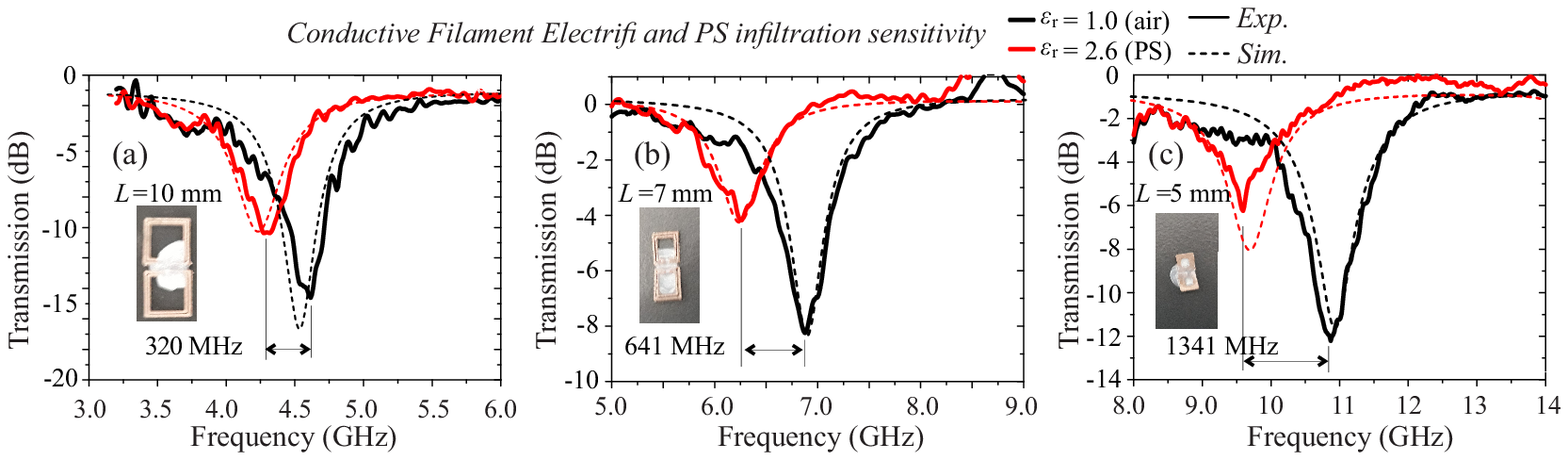}
\caption{\label{fig4}   Tunability of the Electrifi SRRs with respect to the infiltrating Polysterene ($\varepsilon_r$=2.6). Experimental (solid) and numerical (dashed) investigation of the transmission tunability  in the rectangular waveguide for variable  length (a)  $L$=10mm, (b)  $L$=7mm and (c) $L$=5mm.}
\end{figure*}

We now turn to the SRR series made by Electrify filament.  Figure~\ref{fig3}(b) presents the  experimental and numerical characterization of the structures electromagnetic response. In particular we investigate structures with $L = 5,7,10$~mm. Once again, we observe that the unit cell's physical upscaling  leads to a shift of the resonance towards lower frequencies. Additionally, we observe that the resonances are well defined, sharp and relatively deep. Consequently  we see that although the conductivity of the Electrifi Filament ($\sim 1.6\times10^4$ S/m) is orders of magnitude lower than that of cooper, it could be sufficiently high for certain microwave applications.

Finally we evaluate the tunable electromagnetic response of the structure in a presence of an infiltrating material. When an electromagnetic wave interacts with the SRR, alternating currents are excited along the elements while charge is accumulated in the adjacent elements.  Consequently, the SRR behaves as an inductor-capacitor (LC) circuit with a characteristic resonance frequency $f_0 \sim 1 / (2\pi\sqrt{LC})$.  At the resonance a large confinement of the fields is certain areas of the meta-atom takes place and in particular in the area between the gaps, as well as  around them.\cite{Rao2018,Chowdhury2011,Penciu2008} Placing a dielectric material in the high field position tunes the resonance. Thus, we evaluate the tunability of the SRR structures made by Electrify filament.  In particular, in the area of the facing gaps of the SRR structures we add a droplet of polysterene (PS), diluted in toluene as seen in Fig.~\ref{fig4}. After slightly heating, the toluene is evaporated and the PS is solidified into the SRR gap; the droplet spreads around the area of the gaps. The polysterene exhibits a dielectric permittivity of $\varepsilon_r$=2.6.  The experimental and numerical study is presented in Fig.~\ref{fig4}.  It is obvious that, the presence of the PS shifts the resonance to lower frequencies, $\Delta f_{PS}=|f_{PS}-f_{air}|$, which is due to the permittivity difference $\Delta \varepsilon_{PS}=|\varepsilon_{PS}-\varepsilon_{air}|$. The corresponding sensitivity is determined by the relation  $S=  \Delta f_{PS} / \Delta \varepsilon_{PS}$. We present the study of the sample with unit cell with size $L = 5, 7, 10$~mm. In the case of $L$ = 10 mm the experimental shift of the resonance is equal to 245 MHz and the corresponding sensitivity is equal to $S_{L=10 \text{mm}} = 153$ MHz/PU (Permittivity Unit). For the other cases the sensitivity is $S_{L=7 \text{mm}} = 400$ MHz/PU and $S_{L=5 \text{mm}} = 838$ MHz/PU. Although the present components are not  optimized for maximum operation and tunability, it is deducted that the acquired technology may provide a platform for in-house, printed on demand and on the fly devices spanning over a large range of frequencies in the microwave regime. For example the obtained sensitivity values are more than twice as large as others reported in the literature, and refer to corresponding SRR-based structures, dedicated for liquid sensing applications.~\cite{Abdolrazzaghi2017,Velez2017,Salim2018} Other  everyday applications involve filters, de-multplexers, wavefront controllers, shieldings, and many more. Additionally, the  automatic nature of the 3D printing, the shape, size, volume adaptation and fabrication speed of the produced components render the approach a very suitable candidate for future applications in  the Internet of Things realm, in robotics, in intelligent environment, etc.~\cite{Liaskos2015}

In conclusion, we have fabricated metasurface units of conductive SRR configurations  using the FDM 3D printing method. The method is low cost, quick and user- and eco-friendly. As spool materials, two different thermoplastic filaments have been used for printing, PLA and Electrifi. The former is a well–known dielectric polymer and therefore the PLA printed SRRs were coated with conductive silver epoxy to obtain high electrical conductivity. The latter is an electrically conductive, polymer-based nanocomposite.  SRR structures with varying dimensions and infiltrating materials were fabricated and characterized  in the microwave regime using standard rectangular waveguides. All fabricated SRRs exhibited well defined electromagnetic response, which is tunable, with respect to their dimensions and infiltrating materials.The exhibited electromagnetic response is inferior, yet comparable to the electromagnetic response of the traditionally fabricated, PCB-based SRRs, of the same scale.  The 3D printing fabrication of SRR based configurations appears as efficient route in the direction of the massive production of high quality, flexible, components dedicated for microwave applications.


\begin{figure*}
\includegraphics{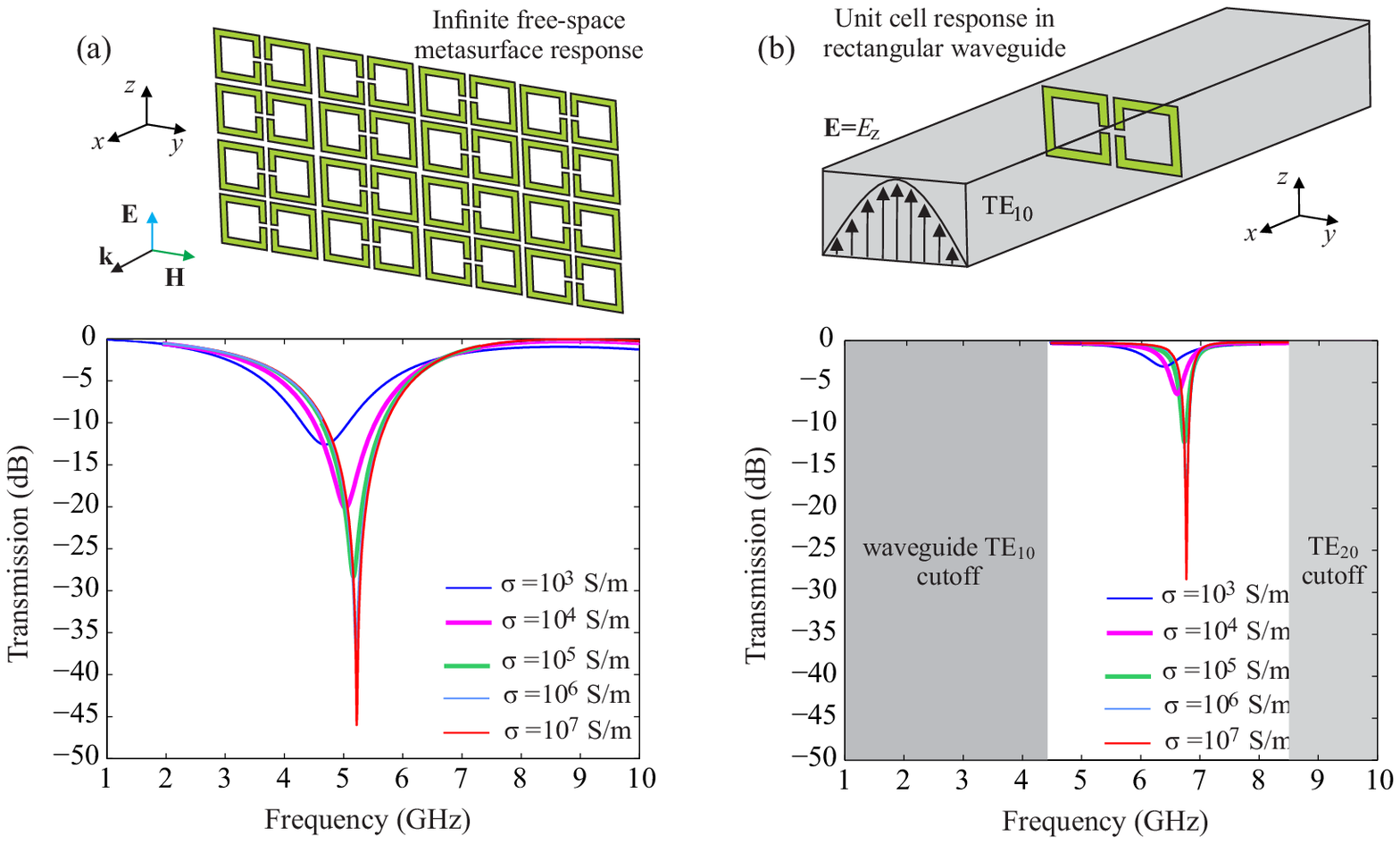}
\caption{\label{wg} Numerical investigation: (a) Transmission spectra from a freestanding free-space, infinite, periodic  metasurface, for various electrical conductivities. The side length of the SRR is $L=6$ mm, the width of the metallic frames $w=0.4$ mm, the thickness of the metal $t=0.4$ mm. The size of the gap is $d=0.2$ mm and the periodicity of the metasurface is $a_y=2(L+w)$ and $a_z = (L+w)$ and along the $z$ and the $y$ axis. The incidence is normal and the polarization is along z axis, \textbf{E}=$E_z$  (b) Corresponding transmission spectra of the metasurface placed in a rectangular waveguide. The dimensions of the waveguide are $a \times b =$ 35 mm $\times$ 15 mm.  }
\end{figure*}

\section{Supplementary Information}

Initially we study the response of millimeter square, opposing gaps SRRs, infinite metasurface as shown in the inset Fig.~S\ref{wg}(a). When an electromagnetic wave of suitable frequency and polarization interacts with the SRR, alternating currents circulating in the wires are excited which gives rise to an effective inductive response.  On the other hand, the gap and the neighboring metallic elements allow the accumulation of opposite changes which forces an effective capacitance. In all, the SRR behaves as an inductor-capacitor (LC) circuit with the  resonance  being a function of the shape of the ring, the size and position of the gaps and the details of the impinging wave. At its resonance, the SRR confines the impinging electromagnetic energy at the volume of the gap or between the adjacent meta-atoms. The structure is designed for operation in the low gigahertz where we perform the experimental characterizations. For the preliminary numerical design the ring is considered to be made of a high conductivity material ($\sigma =10^7$ S/m), whereas no substrate is assumed; that is the metasurface is free-standing. The overall size of the unit cell, that is the periodicity of the metasurface is equal to $a_y = 2(L+w)$ and $a_z=(L+w)$ along the $z$ and the $y$ axis. In the frequency range under consideration E$_z$ polarization obtains a fundamental resonance at 5.2 GHz. In Fig.~S\ref{wg}(a) we also present the dependence of the electromagnetic response in variable wires conductivity values in the range $\sigma =10^3-10^7$  S/m. As observed in Fig. S1(a) (red curves), in the ideal case of $\sigma =10^7$ S/m, which is the case of the metals like copper at the microwave regime (fabricated in PCB boards), the response is sharp and the transmission dip is as low as -45dB. As the conductivity obtains lower values, the response degrades, that is, it becomes less sharp, the quality factor becomes smaller, the minimum of the transmission peak suppresses, while the resonance experiences a shift towards smaller frequencies. However, still in the case of really substandard conductivity like $\sigma =10^3$ S/m the resonance is well defined and quite low (- 12 dB) in the free-space infinite metasurface.

We also characterized numerically the electromagnetic properties of the metasurface by placing a single unit cell in the center of a closed rectangular waveguide, and measuring the scattering parameters as shown in Fig.~S\ref{wg}(b). The rectangular waveguide allows us to trail the resonance of the metasurface in a simple and accessible manner, using a closed system. The approach is valid since the rectangular waveguide mimics the $E_z$ polarized TEM mode of the free space planewave which impinges normally the metasurface and, at the same time, it artificially imposes a periodicity which corresponds to a metasurface. In particular, the waveguide supports a single propagation mode in a frequency regime which depends on its dimension, $a \times b $. In the single mode operation the waveguide supports the propagation of only the fundamental TE$_{10}$ mode (higher order modes decay fast). The single mode frequency range is delimited between the lowest TE$_{10}$ mode cutoff frequency and the upper TE$_{20}$ mode cutoff frequency. The spatial profile of the TE$_{10}$ mode is characterized by no zero crossings, it is polarized in the $z$ axis, $E_z=\sin \pi / a$, and therefore is as similar as possible to the TEM wave. Additionally, the conditions at the metallic faces of the waveguide are: $E_z(y=0,z)= E_z(y=a,z)$ and $E_z(y,z=0) = E_z(y,z=a)$, and mimic the periodic conditions of the free space metasurface with the restriction of the constant waveguide size with forces an effective periodic unit cell.

Comparing the infinite, free-space metasurface, shown in  Fig.~S\ref{wg}(a), and the waveguide simulations shown in Fig.~S\ref{wg}(b), we observe an overall shift of the resonance towards larger frequencies. This is a consequence of the waveguide metallic walls that impose periodic conditions which effectively corresponds to a more sparse metasurface, that is, with larger periodicity. In particular, the periodicity of the infinite, free-space metasurface is equal to $a_y  \times a_z  = 12.8   \times 6.4$ mm and the dimensions of the waveguide are $a \times b$ = 35 mm $\times$ 15 mm. This practically means that the effective intercoupling between the unit cells is weaker in the waveguide case (and stronger in the free space metasurface where the unit cells are close to each other). Apart from the resonant frequency, this has an impact also on the depth of the transmission peak, which is weaker in the rectangular waveguide. We also presented the conductivity dependent simulation results for the rectangular waveguide study. The effect is the same as in the infinite case. For the high conducting case the transmission dip is -30dB and it the case of the low conductivity is as low as -2dB. Also, a small resonance shift with varying conductivity is observed.
\begin{figure*}
\includegraphics{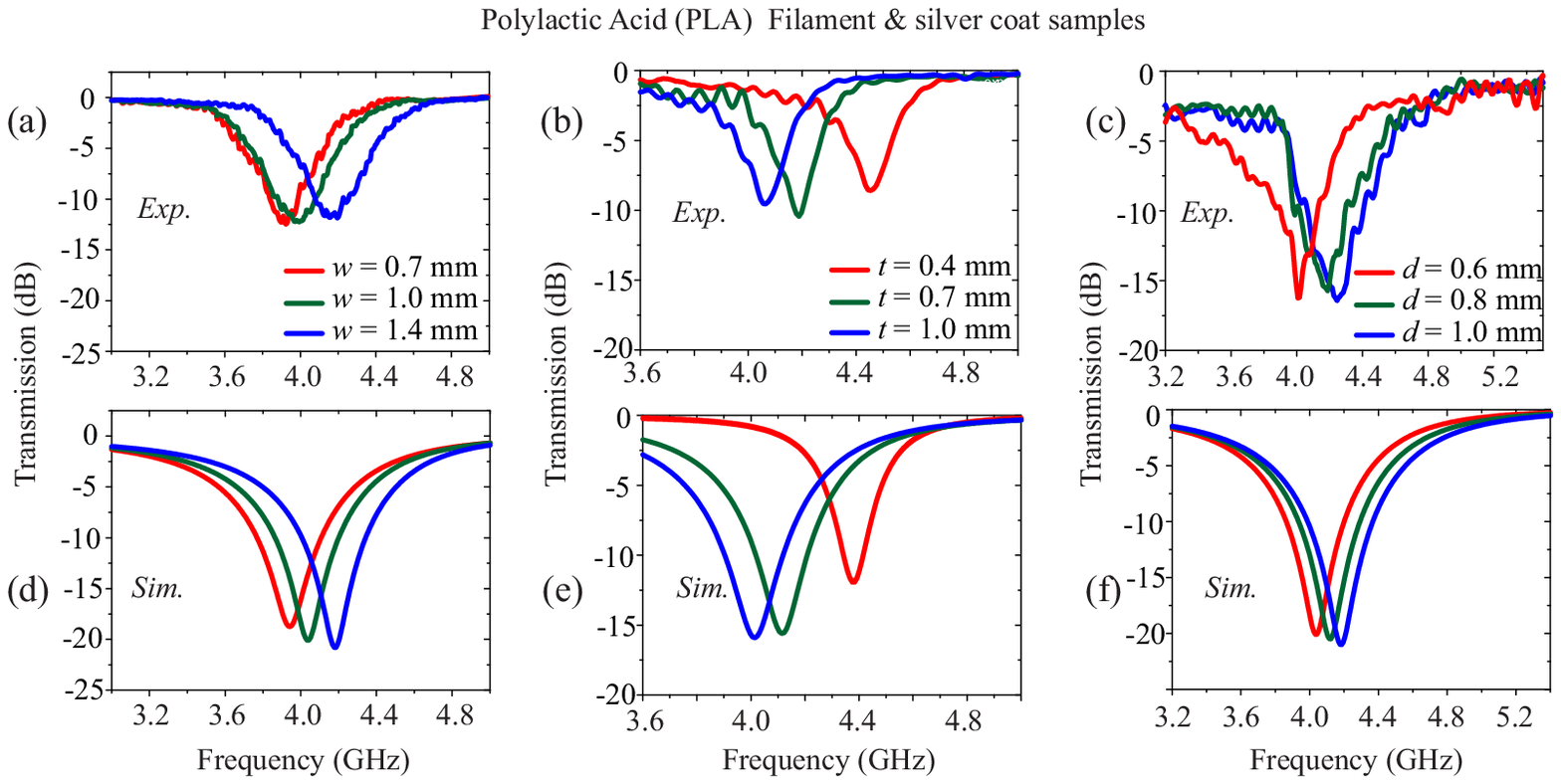}
\caption{\label{vargeom}  (a,d) Transmission spectra for PLA / Ag epoxy coated  SRRs with various widths. Experimental data (a) in comparison with numerical simulations (d). (b,e) Corresponding transmission spectra for SRRs with different wire thickness. Experimental data (b) in comparison with numerical simulations (e).  (c,f) Transmission spectra for SRRs with various gaps. Experimental data (c) in comparison with numerical simulations (f). The side length is equal to $L$ = 10 mm.  }

\end{figure*}

\section{Meta-atom response with variable geometry}
In Fig.~S\ref{vargeom} we study the effect of the  geometrical SRR parameters other than the most determining parameter, the side length $L$ (presented in the manuscript and Fig.~3) on the resonant response of the SRR. The geometrical parameters are found in Fig.~2 of the manuscript. Figure~S\ref{vargeom}(a, d) presents the effect of the SRR width on the transmission resonance (experimental data and numerical data). In particular, we measure the transmission for unit cells with widths$ w = 0.7$ mm (red line), 1 mm (dark green line) and 1.4 mm (blue line), respectively ($L = 10$ mm, $t = 0.4$ mm, $d = 0.6$ mm). The response exhibits only a small dependence on the wire width as the resonance moves towards larger frequencies with increasing $w$. Figure ~S\ref{vargeom}(b, e) presents the effect of the wire thickness in the transmission resonance. In particular, we record the transmission for unit cells with variable thickness, $t = 0.4$ mm (red line), 0.7 mm (dark green) and 1 mm (blue line), respectively ($L = 10$ mm, $w = 2$ mm, $d = 0.6$ mm). The response shows a slight dependence on the wire thickness, as the resonance moves towards smaller frequencies with increasing $t$. Finally, in Fig.~S\ref{vargeom}(c, f) we present the effect of the size of the gap. In particular, we measure the transmission for unit cells with $d = 0.6$ mm (red line), 0.8 mm (dark green line) and 1 mm (blue line) respectively ($L = 10$ mm, $w = 1$ mm, $t = 0.6$ mm). A slight dependence of the resonance frequency on the gap size is observed, as the resonance curve is shifted to higher frequencies, with increasing gap $d$. This is a consequence of the effective decrease of the meta-atom capacitance which leads to a resonance increase. Once again we observe an overall experimental resonance with shallower values than the corresponding numerical ones. It becomes evident that the variation of the length $L$ of the basic SRR component leads to substantial shift of resonance minima, while all the other SRR parameters (width, thickness, gap size) weakly affect the resonance frequency. Thus, adjusting  the dimensions, especially the length of the SRR, we can regulate the operation frequency of the structure.

\section*{Acknowledgment}
This research has been co-financed by the European Union and Greek national funds through the Operational Program Competitiveness, Entrepreneurship and Innovation, under the call RESEARCH–CREATE–INNOVATE (project code: T1EDK-02784; acronym: POLYSHIELD) and by the European Union’s Horizon
2020 FETOPEN programme under project VISORSURF grant agreement No.
736876.

\bibliographystyle{IEEEtran}
\bibliography{shortbib}


\end{document}